\documentclass[aps,prl,superscriptaddress,reprint]{revtex4-1}
\usepackage{centernot}
\usepackage{graphicx}
\usepackage{amsmath}
\usepackage{times}
\usepackage{amssymb}
\usepackage{mathrsfs}
\usepackage{chemarr}
\usepackage{xcolor}
\usepackage{url}
\usepackage{version}
\newcommand\Ft{\mathcal{F}_{\rm topo}}

\graphicspath{{./Image/}}

\usepackage[pdftex,colorlinks=true,pdfstartview=FitV,linkcolor= linkcolor,citecolor= linkcolor,urlcolor= linkcolor,hyperindex=true,hyperfigures=false]{hyperref}
\definecolor{linkcolor}{rgb}{0,0,0.6}

\newcommand\bfu{{\bf u}}

\usepackage{tikz}

\begin{document}


\title{Fluctuation-induced phase separation in metric and topological models of collective motion}

\author{David Martin}
\affiliation{Université de Paris, Laboratoire Mati\`ere et Syst\`emes Complexes (MSC), UMR 7057 CNRS,  F-75205 Paris,  France}
\author{Hugues Chat\'e}
\affiliation{Service de Physique de l'\'Etat Condens\'e, CNRS UMR 3680, CEA-Saclay, 91191 Gif-sur-Yvette, France}
\affiliation{Computational Science Research Center, Beijing 100094, China}
\author{Cesare Nardini}
\affiliation{Service de Physique de l'\'Etat Condens\'e, CNRS UMR 3680, CEA-Saclay, 91191 Gif-sur-Yvette, France}
\author{Alexandre Solon}
\affiliation{Sorbonne Universit\'e, CNRS, Laboratoire Physique Th\'eorique de la Mati\`ere Condens\'ee, 75005 Paris, France }
\author{Julien Tailleur}
\affiliation{Université de Paris, Laboratoire Mati\`ere et Syst\`emes Complexes (MSC), UMR 7057 CNRS,  F-75205 Paris,  France}
\author{Fr\'ed\'eric Van Wijland}
\affiliation{Université de Paris, Laboratoire Mati\`ere et Syst\`emes Complexes (MSC), UMR 7057 CNRS,  F-75205 Paris,  France}
\date{\today}

\begin{abstract}
  \if{We study the onset of polar order in dry active matter at the
    field-theoretical level.  We show that fluctuations generically
    force the transitions predicted by deterministic hydrodynamic
    theories into a phase-separation scenario by inducing a
    density-dependent shift of the onset of collective motion.  Our
    results first apply to metric models whose deterministic
    hydrodynamics spuriously predict second order transitions. They
    also hold for topological models in which particles interact with
    a fixed number of nearest-neighbors, which were believed so far to
    exhibit a continuous onset of order. Our results are confirmed by
    numerical simulations of fluctuating hydrodynamics and microscopic
    models.}\fi
  \if{We study the onset of polar order in dry active matter at the
  field-theoretical level. Starting from field theories that predict a
  continuous transition at the deterministic level, we show that
  fluctuations induce a density-dependent shift of the onset of order,
  which in turns changes the nature of the transition into a
  phase-separation scenario. Our results first apply to metric models
  whose deterministic hydrodynamics spuriously predict second order
  transitions. They also hold for topological models in which
  particles interact with a fixed number of nearest neighbors, which
  were believed so far to exhibit a continuous onset of order. Our
  results are confirmed by numerical simulations of fluctuating
  hydrodynamics and microscopic models.}\fi
We study the role of noise on the nature of the transition to
collective motion in dry active matter. Starting from field theories
that predict a continuous transition at the deterministic level, we
show that fluctuations induce a density-dependent shift of the onset
of order, which in turns changes the nature of the transition into a
phase-separation scenario. Our results apply to a range of systems,
including the topological models in which particles interact with a
fixed number of nearest neighbors, which were believed so far to
exhibit a continuous onset of order. Our analytical predictions are
confirmed by numerical simulations of fluctuating hydrodynamics and
microscopic models.
\end{abstract}

\maketitle

Within active matter studies, the transition to collective motion is a
problem of both historical and paradigmatic value, which has led to a
wealth of
theoretical~\cite{toner1995long,toner2005hydrodynamics,bertin2006boltzmann,mishra2010fluctuations,solon2013revisiting},
numerical~\cite{vicsek1995novel,gregoire2004onset,VicsekFirstOrder}
and experimental works, both on
biological~\cite{buhl2006disorder,Ballerini2008,schaller2010polar,cavagna2017dynamic,bain2019dynamic}
and
synthetic~\cite{narayan2007long,deseigne2010collective,bricard2013emergence}
systems. Thanks to its simplicity, the setting of dry polar flocks, in
which self-propelled particles stochastically and locally align
their velocities, continues to inspire the research of
many~\cite{vicsek1995novel,gregoire2004onset,ginelli2010large,peruani2011traffic,solon2013revisiting,dossetti2015emergence,morin2017distortion,chen2018incompressible,lavergne2019group,mahault2019quantitative,geyer2019freezing,charlesworth2019intrinsically,chen2020moving}.

In \textit{metric} models, particles align with all their neighbors
within a finite distance. At the microscopic level, the nature of the
transition is now well
established~\cite{gregoire2004onset,VicsekFirstOrder}, notably thanks
to the introduction of the active Ising model (AIM) in which
rotational symmetry of the Vicsek model (VM) is explicitly
broken~\cite{solon2013revisiting}. For both the AIM and the VM, the
scenario is that of a phase-separation between a disordered
gas/paramagnetic phase and a polarly ordered liquid/ferromagnetic
phase, with a coexistence region whose properties depend on the
symmetry of the model~\cite{solon2015phase}.

At the continuous level, these results have first been accounted for
using deterministic hydrodynamic theories, which typically couple a
density field $\rho$ and an order parameter
field~\cite{bertin2006boltzmann,bertin2009hydrodynamic,solon2013revisiting,caussin2014emergent,ihle2016chapman}.
The nature of the phase transition can indeed be understood by
considering their dynamics in one spatial dimension, a minimal model
of which is given by
\begin{eqnarray}
  \partial_t \rho &=& D \partial_{xx} \rho - v \partial_x m\label{eq:MFrho}\\
  \partial_t m &=& D \partial_{xx} m - v \partial_x \rho -\mathcal{F}(\rho,m)\;.\label{eq:MFm}
\end{eqnarray}
Here, $m$ is akin to a magnetisation field in a spin system and
$\mathcal{F}(\rho,m)=\alpha m + \gamma \frac{m^{3}}{\rho^{2}}$ is a
Landau term that controls ferromagnetic
alignment~\cite{solon2013revisiting}. Equations~\eqref{eq:MFrho}
and~\eqref{eq:MFm} have been derived for the
AIM~\cite{solon2013revisiting}; many similar hydrodynamic models have
been proposed or derived in one and two
dimensions~\cite{bertin2006boltzmann,bertin2009hydrodynamic,toner2005hydrodynamics,mishra2010fluctuations,bricard2013emergence,solon2013revisiting,ihle2016chapman}. All these models lead to the
same conclusion: the first-order nature of the transition stems from
the density-dependence of the linear term: $\alpha=\alpha(\rho)$. To
continue the ferromagnetic analogy, as detailed in~\cite{supp}, any
density-dependent critical temperature such that $\alpha'(\rho)\neq 0$
leads to a phase-separation transition. The latter is characterized by
two main features: homogeneous ordered profiles are linearly unstable
close to the transition, when $\alpha \lesssim 0$, and this
instability leads to the emergence of inhomogeneous propagating
solutions~\cite{gregoire2004onset,caussin2014emergent,Solon2015PatternFI}. Interestingly,
a number of microscopic metric models have also been described using
continuous descriptions in which, in some scaling limit, $\alpha$ is
independent of the density, thus being compatible with continuous
scenarios~\cite{barbaro2014phase,barre2015motility}.

In a second class of models, refered to as
\textit{topological}~\cite{Camperi2012Interface} and motivated by
animal-behavior studies, the interaction between two particles is not
decided based on their relative distance. For instance, particles can
align with their Voronoi neighbors~\cite{ginelli2010relevance} or with
their $k$ nearest neighbors~\cite{Ballerini2008}. Doubling the
distance between all particles, and hence reducing the particle
density, does not impact the aligning dynamics so that these models
are expected to be less sensitive to density variations. The
coarse-graining of a Voronoi-based model has indeed led to a
density-independent critical temperature, hence predicting a
continuous onset of order~\cite{peshkov2012continuous}. Existing
numerics appear consistent with such a continuous
scenario~\cite{ginelli2010relevance}.

Finally, deterministic hydrodynamic equations only capture part of the
large-scale physics and the role of fluctations at the coarse-grained
level must also be assessed. In the homogeneous ordered phase, this
was done using the celebrated Toner-Tu equations to establish the
existence of long-range orientational order in two
dimensions~\cite{toner1995long,toner2005hydrodynamics,toner2012reanalysis}. For
metric systems with density-dependent critical temperature, the
phase-separation scenario is robust to the introduction of
noise~\cite{mishra2010fluctuations}. The latter, however, selects the
nature of the propagating bands observed in the coexistence
regime~\cite{solon2015phase}. On the contrary, the impact of
fluctuations on the continuous scenario, for both metric and
topological models with density-independent critical temperature,
remains to be explored.

In this Letter, we show how fluctuations generically turn continuous
transitions to collective motion predicted by deterministic
hydrodynamics into the standard phase-separation scenario.  We first
show that dressing the PDEs~\eqref{eq:MFrho}-\eqref{eq:MFm} with noise
generically yields a renormalized density-dependent critical
temperature, hence leading to phase separation. We note that the
hydrodynamic theory of topological interactions introduced
in~\cite{peshkov2012continuous}, which is based on a Boltzmann
approach that successfully captured the nature of the transition in
metric models~\cite{bertin2006boltzmann,peshkov2014boltzmann}, leads
to a description equivalent to \eqref{eq:MFrho}-\eqref{eq:MFm} with a
density-independent $\alpha$. Dressing it with noise will thus also
lead to a phase-separation scenario. As an alternative field theory of
topological models, we propose a modification of the Landau term
in~\eqref{eq:MFm} that preserves the topological nature of the
interaction at the coarse-grained level. Surprisingly, this still
leads to a fluctuation-induced first-order scenario. To probe whether
this prediction is an artifact of our continuous theory or a generic
feature of topological interactions, we consider a microscopic model
in which particles align with their $k$ nearest neighbors. Numerical
simulations confirm a liquid-gas phase separation scenario. Finally, we
show how measuring the dependency of the onset of order on the average
density is a simple quantitative test that allows predicting the
nature of the transition. All calculations below are based on the
one-dimensional hydrodynamic theory~\eqref{eq:MFrho}-\eqref{eq:MFm}
and its topological generalization. Our results can be extended to two
dimensions and to other hydrodynamic models~\cite{DavidFutur}. We
complement our analytical approach by numerical simulations, mostly in
2D, which are all detailed in~\cite{supp}.

\textit{Fluctuation-induced first-order transitions.} We first study
the impact of fluctuations on the deterministic
dynamics~\eqref{eq:MFrho} and~\eqref{eq:MFm}, which encompass metric
models, but also the hydrodynamic theory proposed for Voronoi
neighbors~\cite{peshkov2012continuous}. We consider $\alpha$
independent of $\rho$ to study the fate of the continuous transition
predicted by~\eqref{eq:MFrho}-\eqref{eq:MFm} in this case. To do so,
we complement Eq.~\eqref{eq:MFm} with a noise term:
\begin{equation}\label{eq:SDEmetric}
  \partial_t m = D \partial_{xx} m - v \partial_x \rho -\mathcal{F}(\rho,m) + \sqrt{2 \sigma \rho} \, \eta
  \;,
\end{equation}
where $\eta(x,t)$ is a zero-mean delta-correlated Gaussian white noise
field. Note that, hereafter, $\rho(x,t)$ and $m(x,t)$ represent
fluctuating fields. The order parameter $m(x,t)$ represents the sum of
the orientations of particles located around position $x$. The noise
acting on $m(x,t)$ will thus be multiplicative; it describes the
fluctuations of a sum over $\propto \rho$ particles and we take it
proportional to $\sqrt{\rho(x,t)}$.  We now construct the
hydrodynamics of the average fields $\rho_0(x,t)=\langle
\rho(x,t)\rangle$ and $m_0(x,t)=\langle m(x,t) \rangle$ to leading
order in the noise strength $\sigma$, where brackets represent average
over noise realizations. In principle, we could also complement
Eq.~\eqref{eq:MFrho} with a conserved noise. The latter is expected to
be subdominant at large scales and we ignore it here, although our
approach can be extended to this case. Introducing $\delta \rho = \rho
- \rho_0$ and $\delta m = m - m_0$, the dynamics of $\rho_0$ and $m_0$
can  be approximated as
\begin{eqnarray}
\partial_{t} \rho_0 &=& D\partial_{xx}\rho_0-v\partial_{x}  m_0\label{eq:RenMet_rho} \\
\notag \partial_{t} m_0 &=& D \partial_{xx} m_0 - v\partial_{x}  \rho_0  - {\cal F}(\rho_0,m_0) \\ \label{eq:RenMet_m} &-&  \frac{\partial^{2} \mathcal{F}}{\partial m^{2}} \frac{\langle \delta m^{2}\rangle}2  -  \frac{\partial^{2} \mathcal{F}}{\partial \rho^{2}} \frac{\langle \delta \rho^{2}\rangle}2 - \frac{\partial^{2} \mathcal{F}}{\partial m \partial \rho} \langle \delta m\delta \rho\rangle
\end{eqnarray}
 To close Eqs.~\eqref{eq:RenMet_rho} and~\eqref{eq:RenMet_m}, we need
 to compute $\langle \delta m^{2}\rangle$, $\langle \delta
 \rho^{2}\rangle$, $\langle \delta m\delta\rho\rangle$ as functions of
 $\rho_0$ and $m_0$. In the small noise limit, the fluctuations
 $\delta \rho$ and $\delta m$ are assumed to be small so that we
 compute these correlators at the linear, Gaussian fluctuations
 level~\cite{nazarenko2000nonlinear,marston2008statistics,bouchet2013kinetic,bakas2014theory}. The
 dynamics of $\delta \rho(x,t)$, $\delta m(x,t)$ then read
\begin{eqnarray}
\partial_{t} \delta \rho &=& D \partial_{x}^2\delta\rho-v\partial_{x} \delta m \label{eq:LinFluct_rho} \\
\partial_{t} \delta m\! &=&\! D \partial_{x}^2\delta m\!-\!v\partial_{x} \delta \rho\! -\!
\frac{\partial \mathcal{F}}{\partial \rho}\delta \rho\! -\!
\frac{\partial \mathcal{F}}{\partial m}\delta m\! +\! \sqrt{2 \sigma \rho_0} \eta\
\label{eq:LinFluct_m}
\end{eqnarray}
Away from the deterministic transition $\alpha=0$, this linear system
of equations leads to bounded fluctuations of $\delta\rho$, $\delta m$
around the homogeneous solutions of Eq.~\eqref{eq:MFrho}
and~\eqref{eq:MFm}. It can be solved in Fourier space and
the correlators appearing in~\eqref{eq:RenMet_m} can be obtained
explicitly as integrals over $k$ space, e.g.  $\langle \delta m^2
\rangle = \int {\rm d}k \langle \delta m_k \delta m_{-k}
\rangle/(2\pi)$~\cite{supp}. The alignment terms in
Eq.~\eqref{eq:RenMet_m} can then, consistently with the approximation
leading to a Landau form, be expanded as $\tilde {\cal
  F}(\rho_0,m_0)=\tilde \alpha m_0 + \tilde \gamma
m_0^3/\rho_0^2$. Fluctuations have thus, to this order in $\sigma$,
dressed $\alpha$ and $\gamma$ into $\tilde \alpha$ and
$\tilde\gamma$. For a given set of parameters, the integrals over $k$
space can be computed numerically. It is, however, more enlightening
to compute them explicitly in the high temperature phase, where
$\alpha>0$. The precise expression of $\tilde \gamma$ is irrelevant
for our purpose, and is presented in~\cite{supp}. The linear term is
renormalized into
\begin{eqnarray}
\tilde{\alpha} &=& \alpha + \frac{3 \sigma \gamma}{ 4 \rho_0 v} f\Big(\frac{\alpha D}{v^2}\Big)\;\text{with}\;f(u)=\frac{\sqrt{2/u}+\sqrt{1+u}}{2+u}\,.
\label{eq:RenAlpha}
\end{eqnarray}
Importantly, $\tilde \alpha$ now depends explicitly on the density.
To first order in $\sigma$, fluctuations thus renormalize the
continuous transition predicted by Eqs.~\eqref{eq:MFrho}
and~\eqref{eq:MFm} into the standard liquid-gas phase separation. Note
that higher orders in $\sigma$ have no reason to cancel the dependence
of $\tilde \alpha$ on density and we thus expect our conclusions to
hold non-perturbatively in $\sigma$.

\begin{figure}
  \centering
  \begin{tikzpicture}
    \path (-.9,0) node {\includegraphics[width=.485\columnwidth]{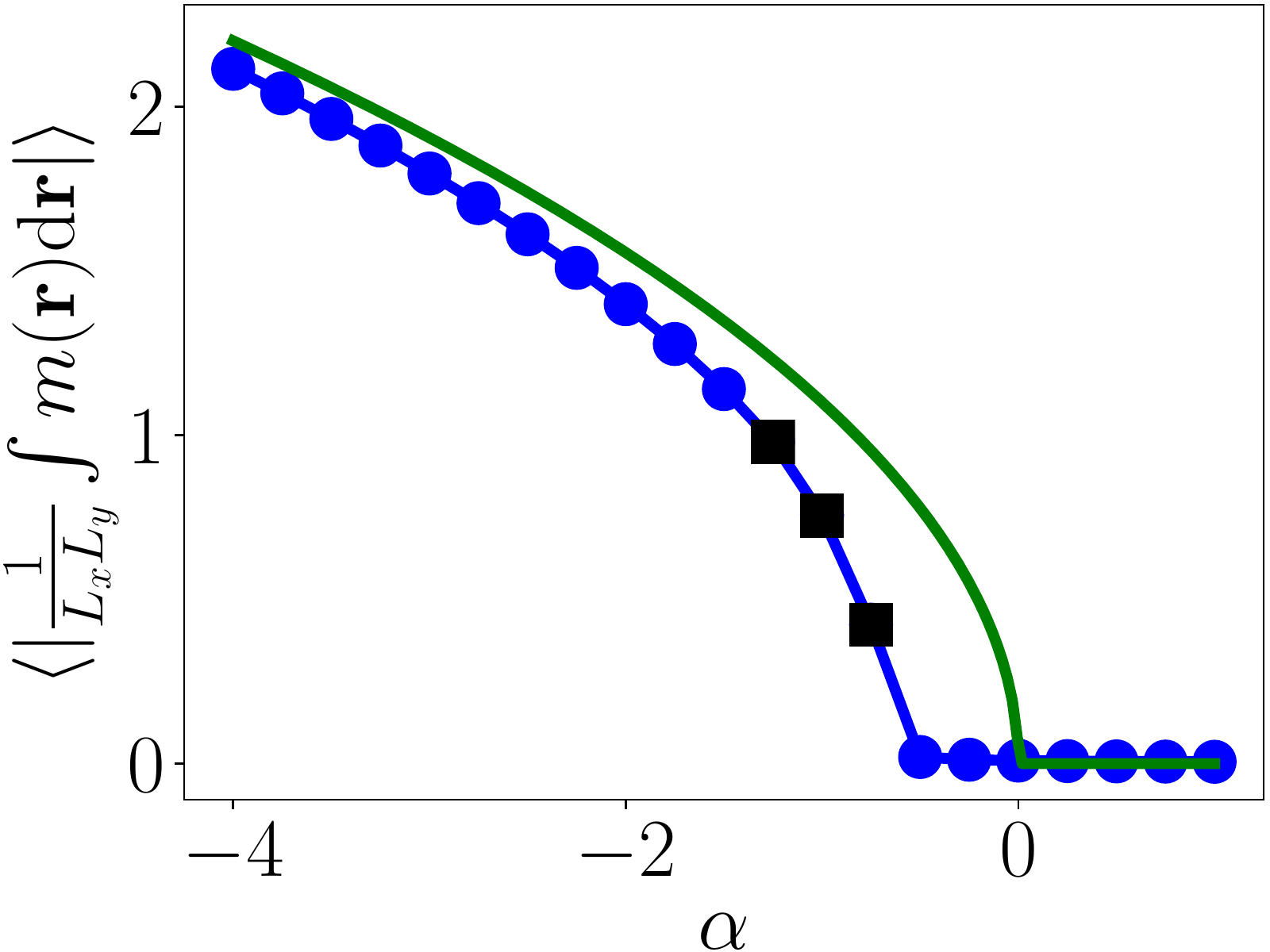}};
    \path (3.4,0) node {\includegraphics[width=.485\columnwidth]{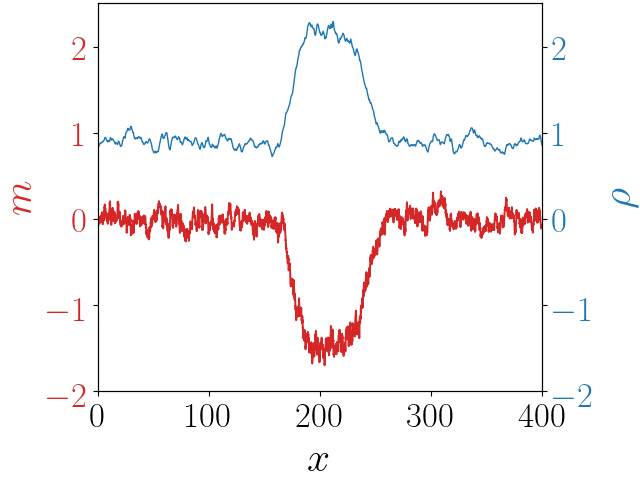}};
    \def\y{-2.75}
    \path (1,\y) node {\includegraphics[width=.95\columnwidth]{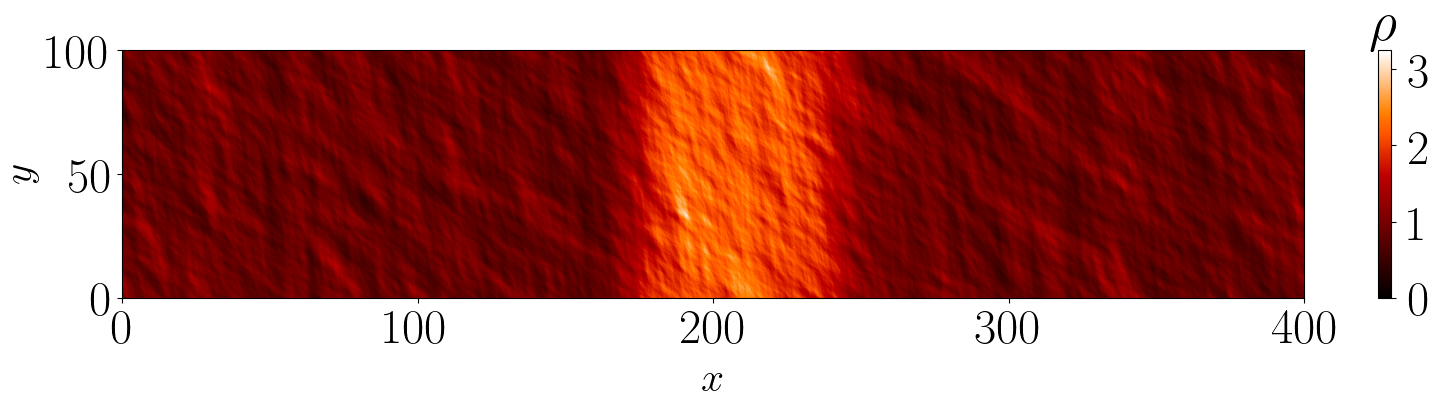}};
    \draw[line width=3pt,blue!50!white,->] (1,\y+.15) -- +(-1.5,0);
  \end{tikzpicture}
  \caption{Simulations of the 2D generalization of
      Eqs.~\eqref{eq:MFrho} and~\eqref{eq:SDEmetric} detailed
      in~\cite{supp}. {\bf Top left}: Average magnetisation as
      $\alpha$ is varied. The transition occurs at $\alpha_c<0$,
      shifted from the mean-field prediction (green line). At the
      onset of order, inhomogeneous profiles (black squares) separate
      homogeneous ordered and disordered phases (blue
      dots). Parameters: $D=v=\gamma=\sigma=1$, $dx=0.5$, $dt=0.01$,
      $L_x=400$, $L_y=40$, $\bar\rho\equiv N/(L_x L_y)=1.1$. {\bf
        Bottom}: A snapshot close to the transition shows ordered
      travelling bands in a disordered background.  {\bf Top right}:
      The corresponding density and magnetization fields averaged
      along $y$. Parameters: same as before up to $L_y =
      100$, $dx=0.1$, $\alpha = -0.9$. }
    \label{fig:SDEmetric}
\end{figure}

To confirm our predictions, we carried out simulations of the 2D
generalization of the stochastic PDEs~\eqref{eq:MFrho}
and~\eqref{eq:SDEmetric}~\cite{supp}. As shown in
Fig.~\ref{fig:SDEmetric}, the continuous transition predicted
by~\eqref{eq:MFrho} and~\eqref{eq:MFm} is replaced by the standard
liquid-gas
framework~\cite{solon2013revisiting,solon2015flocking}. This is best
illustrated by the emergence of a coexistence region in which ordered
bands travel in a disordered
background~\cite{gregoire2004onset,bertin2006boltzmann,bertin2009hydrodynamic,mishra2010fluctuations}.

\vspace{-0.10cm}

\textit{Field theory for topological
  interactions.} \label{sec::topological_model} The study of the
dynamics~\eqref{eq:MFrho} and~\eqref{eq:SDEmetric} thus showed that
fluctuations generically make the transition to collective motion
first order in metric models. This applies, in particular, to the
hydrodynamic theory proposed for Voronoi-based interactions
in~\cite{peshkov2012continuous}. We now propose an alternative
hydrodynamic description which preserves the topological nature of the
interactions at the coarse-grained level. To do so, we focus on models
in which particles align with their $k$ nearest neighbors.  We
introduce a coarse-grained field $y(x)$ which measures the interaction
range of a particle at $x$:
\begin{equation}\label{eq:definitiony}
  \int^{x+y(x)}_{x-y(x)}\rho(z){\rm d}z=k\;.
\end{equation}
Particles at position $x$ then align with a `topological' field $\bar
m(x,t)$ computed over their $k$ nearest neighbours through
\begin{equation}
  \bar{m}(x)=\frac 1 k \int^{x+y(x)}_{x-y(x)}m(z){\rm d}z\;.
\end{equation}
Doubling the distance between particles does not alter the values of
$\bar m(x)$, consistent with microscopic topological
models~\cite{Ballerini2008,ginelli2010relevance}. To construct the
topological counterpart of the Landau term $\mathcal{F}(\rho,m)$
appearing in the metric dynamics, let us recall how the latter is
constructed from microscopic models. In a ferromagnetic context,
$\mathcal{F}$ can be seen as the small-magnetisation expansion of a
more complex function $\mathcal{F}_{\rm ferro.}=2m\cosh(\beta
p)-2\rho\sinh(\beta p)$, where $p=m/\rho$ is the local magnetisation
per particle and $\beta$ the inverse temperature. The fields $\rho$
and $m$ enter $\mathcal{F}_{\rm ferro.}$ through counting statistics,
$(\rho \pm m)/2$ representing the local densities of particles with
plus or minus spins. The field $p$, on the other hand, enters via
the aligning rate at which a spins $s$ flips, e.g.  $W(s\to -s)=
\Gamma \exp(-\beta s p)$. When particles align stochastically with a
topological field $\bar m$, the Landau term thus simply becomes
$\mathcal{F}_{\rm ferro.}=2m\cosh(\beta \bar m)-2\rho\sinh(\beta \bar
m)$. Expanding to third order in the fields then yields:
\begin{equation}
\label{eq:landau_topol}
 \Ft\left(m,\rho,\beta\right) = \Gamma\big(2 m-2 \rho\beta \bar{m}
-\frac{ \rho\beta^3}{3} \bar m^3 +  \beta^2 m \bar m^2\big)
\end{equation}
in which, for simplicity, we retain $\beta$ as the sole control parameter.
At mean-field level, our topological field theory is thus given by Eq.~\eqref{eq:MFrho} and~\eqref{eq:MFm}, with $\mathcal{F}$ replaced by $\Ft$.

Homogeneous solutions $\rho_0,m_0$ correspond to $y(x)=k/(2\rho_0)$
and $\bar m= m_0/\rho_0$. The linear term in $\Ft{}$ then reduces to
$2 \Gamma (1-\beta) m_0$, leading to a density-independent
transition at $\beta_m=1$. Linear stability analysis of the
homogeneous solutions then shows that disordered and ordered solutions
are linearly stable for $\beta<\beta_m$ and $\beta>\beta_m$,
respectively~\cite{supp}. Our topological field theory thus predicts a
continuous transition at the mean-field level.

Let us now assess the effect of dressing the dynamics of the order
parameter with noise: we consider the stochastic
dynamics~\eqref{eq:MFrho} and~\eqref{eq:SDEmetric}, albeit with
$\mathcal{F}$ replaced by $\Ft$. Here also, $m(x,t)$ is the sum of the
orientations of particles located around position $x$ and we consider
a multiplicative noise proportional to $\sqrt{\rho(x,t)}$. To
construct the dynamics of the average fields $\rho_0$ and $m_0$
perturbatively in $\sigma$, we first stress that
Eq.~\eqref{eq:definitiony} directly enslaves the field $y(x)$ to
$\rho(x)$. There are thus, again, only two independent fields,
$\rho(x,t)$ and $m(x,t)$, so that Eq.~\eqref{eq:LinFluct_m} is still
valid, up to $\mathcal{F} \to \Ft$. The expression of $\Ft$ being,
however, more complicated than in the metric case, the algebra is
correspondingly more involved. We detail in~\cite{supp} the
renormalization of the linear part of $\Ft$, which controls the nature
of the transition. To first order in the noise strength $\sigma$, we
find
\begin{equation}
\label{eq::m_topol_ising_fluct_mean}
  (\beta-1) m_0 \to  \left[\beta-1-\sigma\frac{ g\left(\beta,\frac{\Gamma k}{v\rho_0},\frac{\Gamma D}{v^2}\right)}{k}\right]m_0\;,
\end{equation}
with $g(\beta,\frac{\Gamma k}{v\rho_0},\frac{\Gamma D}{v^2})$ a positive function whose expression is provided
as an integral in~\cite{supp}.

Importantly, the linear term in the aligning dynamics of $m_0$ has
again become density-dependent, hence predicting a phase-separation
scenario. This is confirmed by numerical simulations
of~\eqref{eq:MFrho} and~\eqref{eq:SDEmetric} with $\mathcal{F}\to\Ft$,
which again reveal the existence of inhomogeneous propagating
bands~\cite{supp}. Our results thus also predict a fluctuation-induced
phase-separation scenario for topological models.
\begin{figure}
\includegraphics[width=.495\columnwidth]{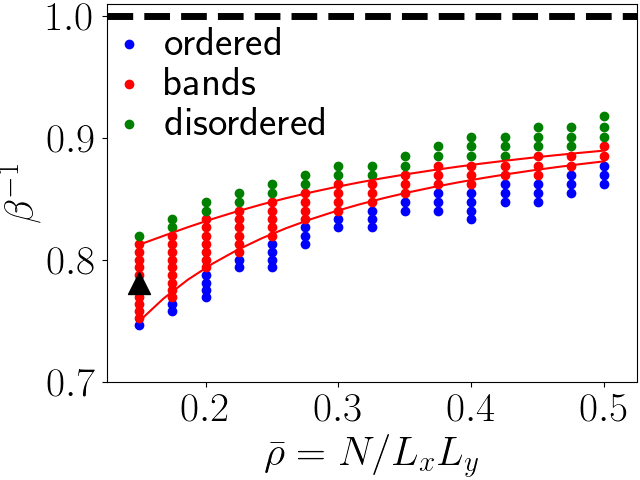}\hspace{-.01\columnwidth}
\includegraphics[width=.5\columnwidth]{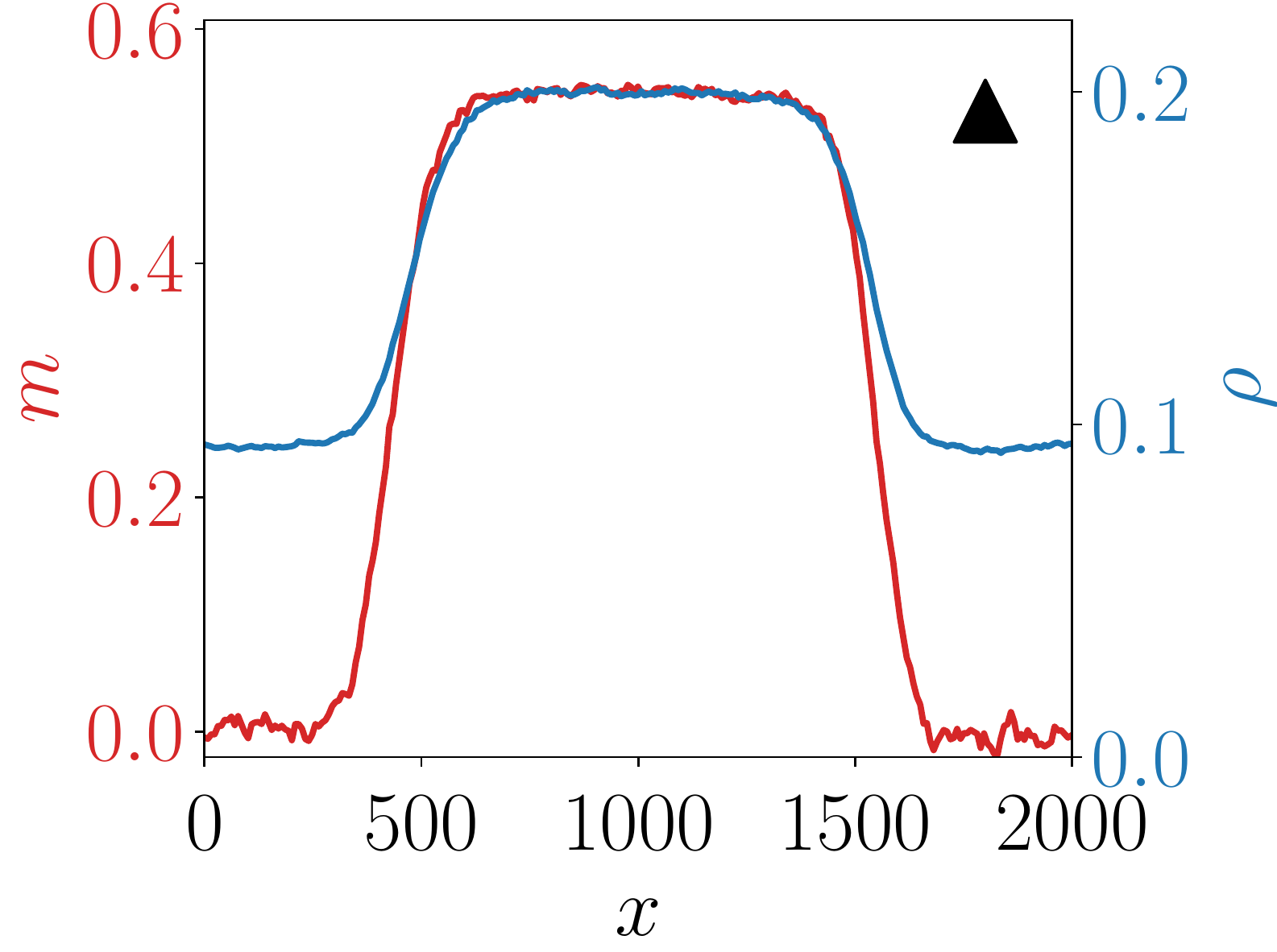}
\hspace{.3cm}\includegraphics[width=.9\columnwidth]{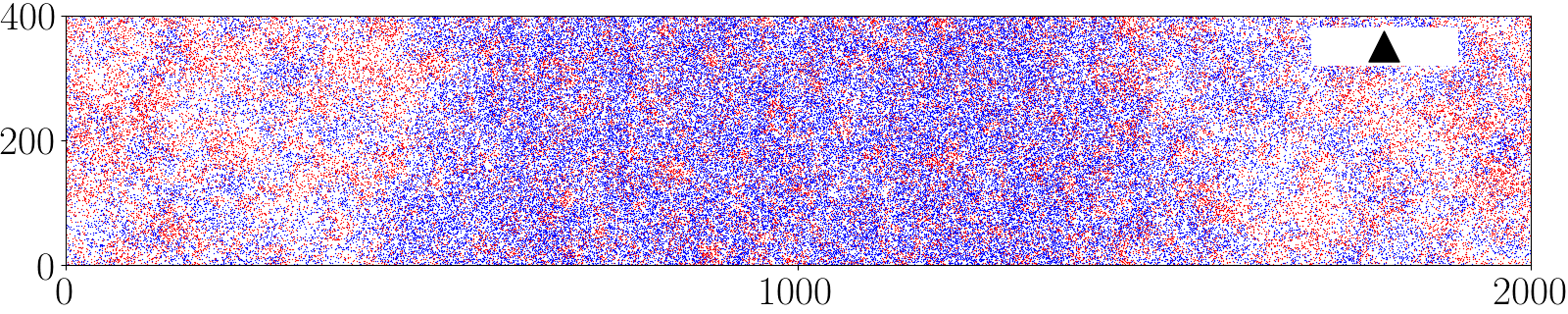}
\caption{{\bf Top Left:} Phase diagram of the microscopic topological
  model defined in Eqs.~\eqref{eq::langevin_spin}
  and~\eqref{eq::flip_dynamic}. The homogeneous ordered (blue) and
  disordered (green) regions are separated by a coexistence phase
  (red). The mean-field critical temperature is $\beta=1$
  (dashed-lined). The red lines are guide to the eyes which show how
  the transitions shift as the mean density varies.  {\bf Bottom:}
  Snapshot of a propagating band corresponding to the black triangle
  in the phase diagram. Blue and red particles correspond to positive
  and negative spins. The corresponding density and magnetization
  fields, averaged over $y$ and time, are shown in the top-right
  panel.  Parameters: $D=8$, $L_{y}=400$, $L_{x}=2000$, $k=3$, $\Gamma
  =0.5$, $v=0.9$.}
\label{fig:phase_diagram}
\end{figure}

\textit{Microscopic models with $k$-nearest-neighbors interactions.}
\label{sec::model}
To test the above predictions, we first consider an off-lattice active
Ising model~\cite{solon2013revisiting,solon2015flocking} in which $N$
particles move in an $L_{x}\times L_{y}$ domain with periodic boundary
conditions. Each particle carries a spin $s_i=\pm 1$ and evolves
according to the Langevin dynamics
\begin{equation}
\label{eq::langevin_spin}
\dot{\textbf{r}_{i}} = s_{i}\ v_0\ \textbf{u}_{x} +\sqrt{2D}\boldsymbol{\eta}_{i}\;,
\end{equation}
where $v_0$ sets the self-propulsion speed and $D$ sets the strength of the Gaussian white noise $\boldsymbol\eta_i$. Spins flip from $s_i$ to $-s_i$ at rate $W(s_i)$ given by
\begin{equation}
\label{eq::flip_dynamic}
W(s_i) = \Gamma e^{-\beta  s_{i}\bar m_i},\quad\text{where}\quad  \bar m_i=\frac 1 k\sum_{j\in\mathcal{N}_{i}} s_{j}\;,
\end{equation}
where $\mathcal{N}_{i}$ is the set of the $k$-nearest neighbours of
particle $i$ and $\bar m_i$ their average magnetization. Note that a
mean-field treatment of the aligning dynamics~\eqref{eq::flip_dynamic}
indeed leads to Eq.~\eqref{eq:landau_topol}.  In agreement with our
predictions, the model exhibits a first order transition to collective
motion, akin to a liquid-gas phase separation: the onset of order at
$\beta\gtrsim \beta_c(\rho_0)$ occurs through the emergence of an
ordered propagating band (Fig.~\ref{fig:phase_diagram}). Unlike the
mean-field critical temperature, the boundaries of the coexistence
region show a clear dependence on the mean density $\bar \rho$.

\begin{figure}
\centering
\begin{tikzpicture}
\path (0,1) node {\includegraphics[totalheight=1.50cm]{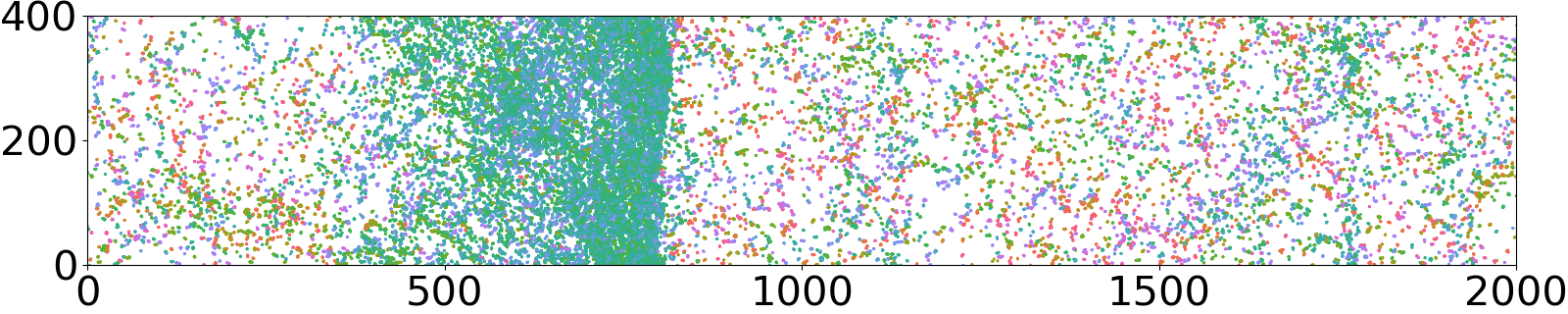}};
\path (0,-1) node {\includegraphics[totalheight=1.50cm]{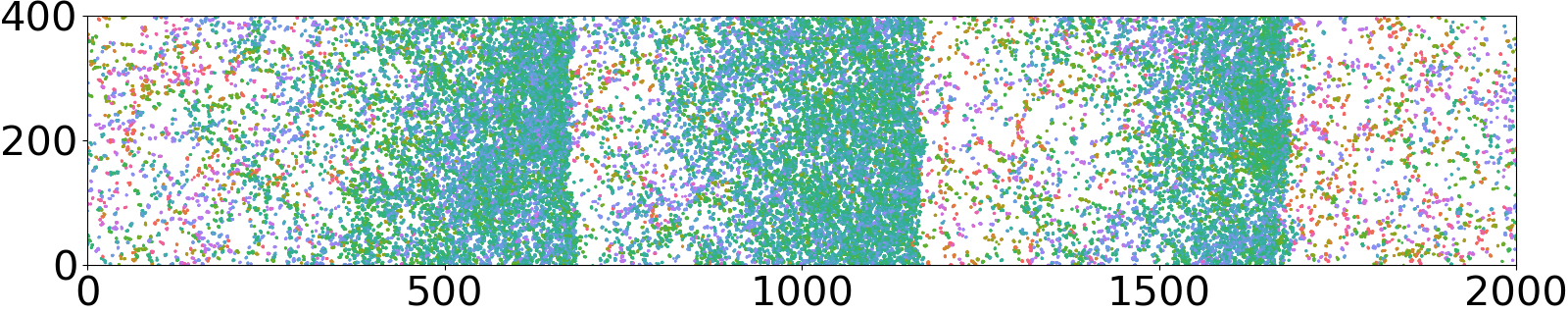}};
\path (4.3,0) node {\includegraphics[totalheight=1.15cm]{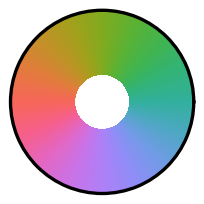}};
\path (4.3,0) node {$\theta$};
\end{tikzpicture}
\caption{Simulations of the topological Vicsek model in 2D. At small
  noise, the system is disordered at low enough densities. Increasing
  the density then leads to an onset of order accompanied by
   propagating bands. Particles align with their $k=3$
  nearest neighbors. Parameters : $L_x = 2000$, $L_y=400$,
  $\sigma=0.08$, $k=3$, $v_0=0.2$, $\Delta t=1$, $\bar \rho=0.25$
  (top) and $\bar\rho=0.4$ (bottom).}
    \label{fig:Vicsektopo}
\end{figure}

To probe the generality of our results, we then implemented a
topological version of the Vicsek
model~\cite{vicsek1995novel,gregoire2004onset} in which the particles
align with their $k$ nearest neighbors. We considered $N$ point
particles carrying unit propulsion vectors $\bfu_i$ and moving in a
$L_{x}\times L_{y}$ domain with periodic boundary conditions. At every
time step, the particles align with the average direction of their $k$
nearest neighbors:
\begin{equation}
  \text{arg}\Big[\bfu_i\Big] \to \text{arg}\Big[\frac{1}{k}\sum_{j\in \mathcal{N}_{i}} \bfu_j \Big] + \sigma \eta\;,
\end{equation}
where $\eta$ is uniformly drawn in $[-\pi,\pi]$.  The particles then
move a distance $v_0\Delta t$ along their propulsion vector. Once
again, propagating bands are observed close to the onset of collective
motion (Fig.~\eqref{fig:Vicsektopo}).

Overall, despite being usually considered resilient to density
fluctuations, the topological interactions studied in this article
thus lead to a phase-separation scenario. Our results suggest that the
nature of the transition to collective motion can be simply assessed
by estimating whether its threshold depends on density or not. We
illustrate this by considering simple models in which the aligning
dynamics is disconnected from spatial positions, hence ensuring that
the transition remains continuous.

\begin{figure}[t!]
\begin{tikzpicture}
\path (0,0) node {\includegraphics[totalheight=2.5cm]{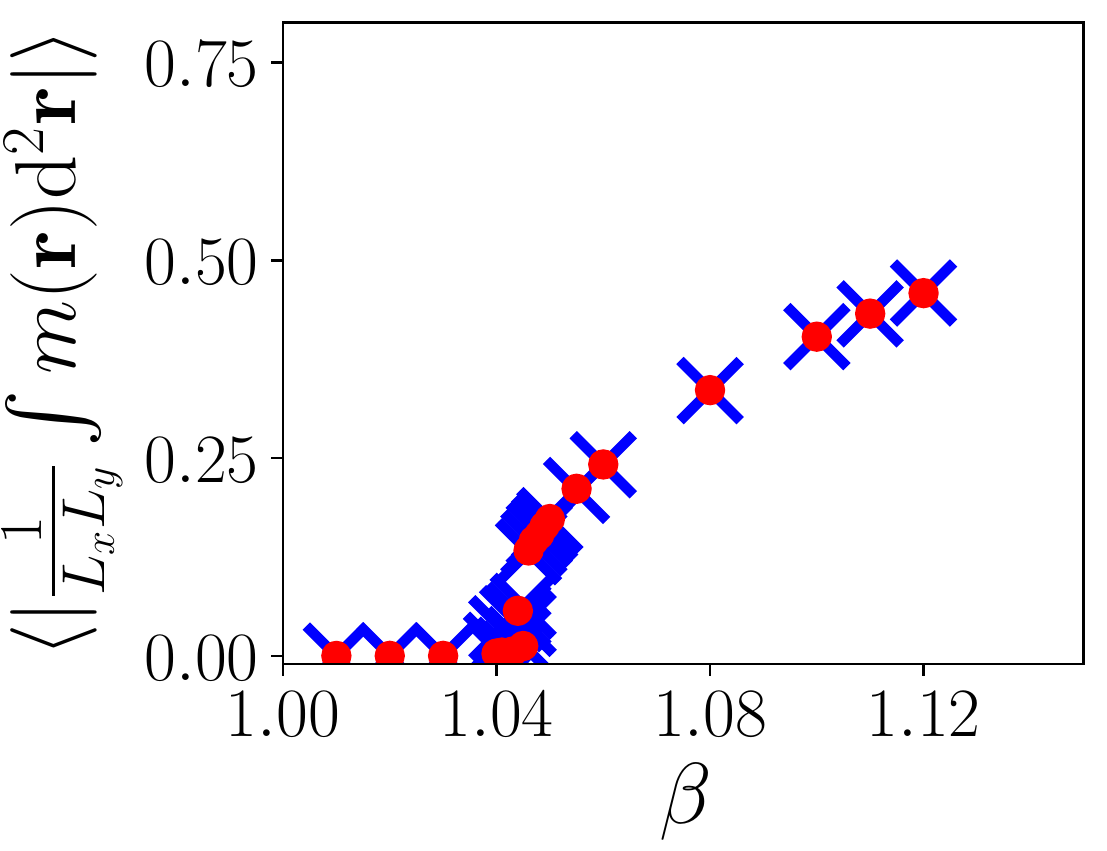}};
\path (3,0) node {\includegraphics[totalheight=2.5cm]{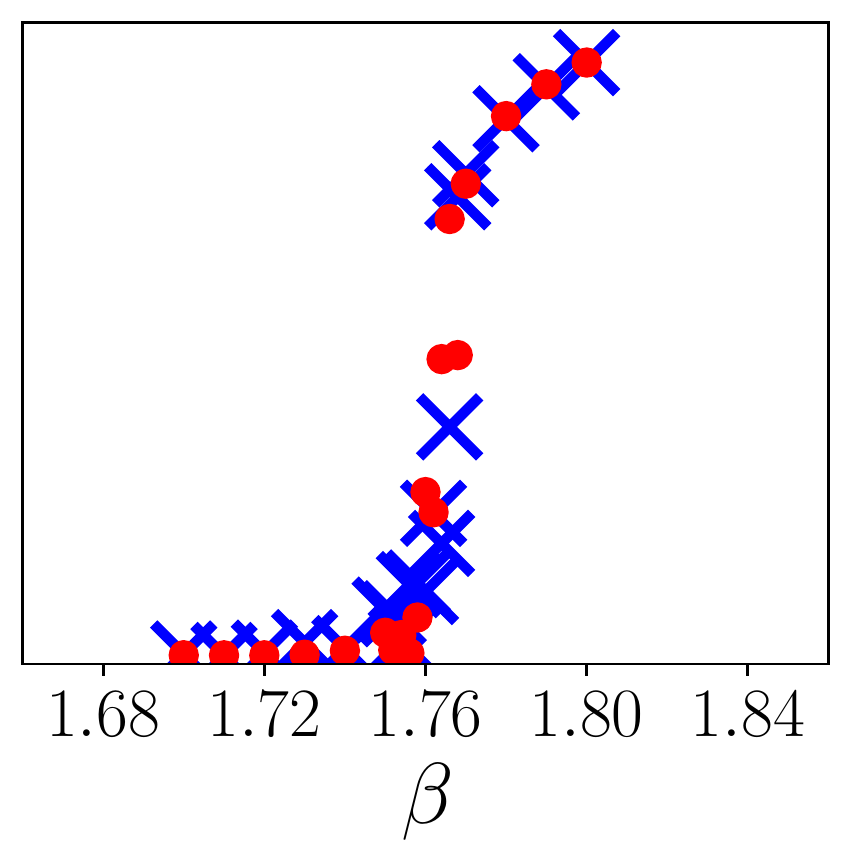}};
\path (5.6,0) node {\includegraphics[totalheight=2.5cm]{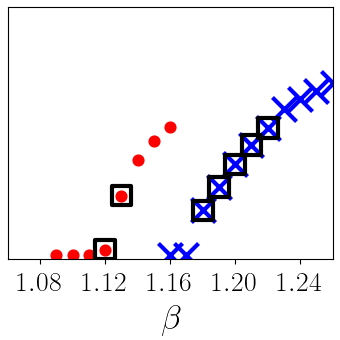}};
\end{tikzpicture}
\caption{Magnetization vs inverse temperature $\beta$ for the RAM
  (left), the LAM (center), and the topological
  model~\eqref{eq::langevin_spin}-\eqref{eq::flip_dynamic}
  (right). Only the latter exhibits travelling bands (black
  squares). Blue crosses and red dots correspond to mean densities
  $\bar \rho=0.25$ and $\bar \rho=0.5$, respectively. Parameters: $L_x
  = 2000$, $L_y = 400$, $D=8$, $v=0.9$, $\Gamma = 0.5$, $k=3$. For
  LAM, $k=4$.}
\label{fig::phase_transition_robust}
\end{figure}

\textit{Loyal and random aligning models.}  We consider $N$ scalar
spins moving in a $L_{x}\times L_{y}$ domain with periodic boundary
conditions according to the Langevin
equation~\eqref{eq::langevin_spin} with two different aligning
dynamics~\footnote{See~\cite{aldana2007phase} for related, though
  different, models.}. In the Random Alignment Model (RAM), the
aligning dynamics is given by~\eqref{eq::flip_dynamic}, with $\bar
m_i$ computed over $k$ spins chosen at random at every time step. In
the Loyal Alignment Model (LAM), on the contrary, alignment occurs
with the same set of $k$ neighbours throughout the simulations,
irrespective of the particle positions. In our simulations, we chose
$k=4$ and assigned to each particle its nearest neighbours on an
initial square lattice. Simulations of both systems lead to
continuous transitions, without the emergence of inhomogeneous phases.
Figure~\ref{fig::phase_transition_robust} shows that measurements of
the global magnetization as the temperature is varied leads to
behaviors which are hard to distinguish between LAM, RAM and our
topological microscopic model. Repeating these measurements at
different densities reveals a density-dependence of the onset of order
in the latter case, but not in LAM \& RAM. Measuring $\beta_c$ as
$\bar\rho$ varies thus constitutes a simple and robust test of the
nature of the transition.

\textit{Conclusion.}  We have shown that dressing hydrodynamic
equations for polar flocks with noise generically leads to a
density-dependent renormalization of the onset of order. Noise thus
makes homogeneous ordered profiles linearly unstable close to the
transition and leads to a phase-separation scenario. This surprisingly
holds for metric-free topological interactions. We confirmed our
field-theoretical computations by numerical simulations of microscopic
models in which particles interact with their $k$ nearest
neighbors. Finally, we have argued that measuring the dependence of
the onset on the global density allows assessing the nature of the
transition. This is likely to be a more stringent test than measuring
putative critical exponents, which are unlikely to distinguish
weakly-first-order transitions from the liquid-gas scenario. Finally,
it would be interesting to reproduce our studies on related models in
which the coupling between density and alignment is qualitatively
altered, from Malthusian~\cite{chen2020moving} to
incompressible~\cite{chen2015critical} or L\'evy
flocks~\cite{cairoli2019active}.

\textit{Acknowledgment:} JT, HC and DM acknowledge support from the ANR grant Bactterns. We thank John Toner for interesting discussions.
CN acknowledges the support of an Aide Investissements d’Avenir du LabEx PALM (ANR-10-LABX-0039-PALM).

\bibliographystyle{apsrev4-1}
\bibliography{biblio_draft_topological}

\end{document}